\documentstyle[prd,aps,preprint,psfig]{revtex}
\begin{document}
\tighten
\title{Double Inflation in Supergravity and the Large Scale Structure}
\author{Toshiyuki Kanazawa} \address{Department of Physics, University
  of Tokyo, Tokyo, 113-0033, Japan} 
\author{M. Kawasaki} \address{Research Center for the Early Universe
  (RESCEU), University of Tokyo, Tokyo, 113-0033, Japan}
\author{Naoshi Sugiyama} \address{Department of Physics, Kyoto
  University, Kyoto, 606-8502} 
\author{T. Yanagida} \address{Department of Physics and RESCEU,
  University of Tokyo, Tokyo 113-0033, Japan}
\date{\today}

\maketitle

\begin{abstract}

  The cosmological implication of a double inflation model with hybrid
  $+$ new inflations in supergravity is studied.  The hybrid inflation
  drives an inflaton for new inflation close to the origin through
  supergravity effects and new inflation naturally occurs.  If the
  total $e$-fold number of new inflation is smaller than $\sim 60$,
  both inflations produce cosmologically relevant density
  fluctuations.  Both cluster abundances and galaxy distributions
  provide strong constraints on the parameters in the double inflation
  model assuming $\Omega_0=1$ standard cold dark matter scenario.  The
  future satellite experiments to measure the angular power spectrum
  of the cosmic microwave background will make a precise determination
  of the model parameters possible.

\end{abstract}



\section{Introduction}

The idea of an inflationary universe~\cite{Guth,Sato} is very
attractive since it can solve serious problems in standard big bang
cosmology such as the horizon and flatness problems~\cite{Guth}.
Though many types of inflation models have been
proposed~\cite{Linde-text}, there are basically three viable models:
chaotic~\cite{Chaotic-inflation}, new~\cite{New-inflation}, and hybrid
inflation~\cite{Hybrid-inflation}. These three models have their own
characters. The chaotic inflation model is difficult to realize in the
framework of supergravity since it requires a classical value of the
inflaton field larger than the gravitational scale ($= 2.4 \times
10^{18}$ GeV and it is taken to be unity throughout this paper). In
supergravity the reheating temperature of inflation should be low
enough to avoid overproduction of gravitinos~\cite{Ellis,Moroi}.  The
new inflation model~\cite{New-inflation} generally predicts a very low
reheating temperature and hence it is the most attractive among the
many inflation models.  However, new inflation suffers from a
fine-tuning problem about the initial condition; i.e., for successful
new inflation, the initial value of the inflaton should be very close
to the local maximum of the potential in a large region whose size is
much longer than the horizon of the universe. On the other hand,
hybrid inflation (and also the chaotic one) can occur for a large
range of initial values.

Recently a framework of double inflation was proposed as a way to
solve the initial value problem of the new inflation
model~\cite{dynamical-tuning}. It was shown that the above serious
problem is solved by supergravity effects if there existed
preinflation (e.g., hybrid inflation) with a sufficiently large Hubble
parameter before new inflation~\cite{dynamical-tuning}.  Different
models of double inflation were studied by various
authors~\cite{doubleinf}. Unlike other double inflation scenarios,
however, our double inflation is quite natural when we try to solve
both the gravitino problem and the initial condition problem in the
new inflation model.

In this double inflation model, if the $e$-fold number of the new
inflation model is smaller than $\sim 60$, density fluctuations
produced by both inflations are cosmologically relevant (the total
$e$-fold number $\sim 60$ is required to solve flatness and horizon
problems in standard big bang cosmology~\cite{Kolb-Turner}). In this
case, the preinflation should account for the density fluctuations on
large cosmological scales [including the cosmic background explorer
(COBE) scales], while the new inflation model produces density
fluctuations on smaller scales.  Although the amplitude of the
fluctuations on large scales should be normalized to the COBE
data~\cite{COBE}, fluctuations on small scales are free from the COBE
normalization and can have arbitrary power matched to the observation.
In Refs.~\cite{KSY-PBH,KY-PBH}, the production of primordial black
hole massive compact halo objects was considered in this double
inflation model. In particular in Ref.~\cite{KY-PBH}, the coherent
oscillation of the inflaton after a preinflation was taken into
account.

In this paper, we study the cosmological implication of the double
inflation model which induces a break on the cosmological ($\gtrsim
$Mpc) scale in the initial density perturbations.  It is well known
that the observations of galaxy distributions cannot be accounted for
with the cosmological density parameter $\Omega_0=1$ and the Hubble
parameter $H_0 = 50$kms$^{-1}$Mpc$^{-1}$ in a standard cold dark
matter (CDM) model.  However, in a double inflation case, there would
be a possibility that the observations may be fit with $\Omega_0=1$
and without a cosmological constant,~\footnote{ Recently it was
  reported that the observations of a supernova type Ia suggests there
  is a nonzero cosmological constant ($\lambda_0$). However, there are
  some papers which point out the problems of interpreting those
  observations~\cite{SN-problems}.} since the produced density
fluctuations would have a nontrivial shape.  Rather we have a chance
to determine parameters of double inflation by observations of the
large scale structure of the universe.  Taking hybrid inflation in
supergravity~\cite{hybrid} as an example of the preinflation, we find
that the produced density fluctuations may account for the observed
clusters abundances~\cite{Eke,Viana-Liddle} and galaxy
distributions~\cite{da-Costa,Lin,Fisher,Feldman,Tadros} with $\Omega_0
=1$ and $H_0 = 50$kms$^{-1}$Mpc$^{-1}$.

\section{Double Inflation Model}

We adopt the double inflation model proposed in
Refs.\cite{dynamical-tuning,KSY-PBH}. The model consists of two
inflationary stages; the first one is called preinflation and we adopt
hybrid inflation~\cite{hybrid} as the preinflation. We also assume
that the second inflationary stage is realized by a new inflation
model~\cite{Izawa-New-inflation} and its $e$-fold number is smaller
than $\sim 60$.  Thus, the density fluctuations on large scales are
produced during the preinflation and their amplitude should be
normalized to the COBE data~\cite{COBE}. On the other hand, the new
inflation model produces fluctuations on small scales. Since the
amplitude of the small scale fluctuations is free from the COBE
normalization, we expect that the new inflation model can produce
density fluctuations appropriate for the observations.

\subsection{Preinflation}

First, let us discuss the hybrid inflation model which we adopt to
cause the preinflation.  The hybrid inflation model contains two kinds
of superfields: one is $S(x,\theta)$ and the others are
$\Psi(x,\theta)$ and $\bar{\Psi}(x,\theta)$. Here $\theta$ is the
Grassmann number denoting superspace. The model is based on the
U$(1)_R$ symmetry under which $S(\theta) \rightarrow e^{2i\alpha}
S(\theta e^{-i\alpha})$ and $\Psi(\theta) \bar{\Psi}(\theta)
\rightarrow \Psi(\theta e^{-i\alpha}) \bar{\Psi}(\theta
e^{-i\alpha})$. The superpotential is given
by~\cite{Hybrid-inflation,hybrid}
\begin{equation}
    W(S,\Psi,\bar{\Psi}) = -\mu^{2} S + \lambda S \bar{\Psi}\Psi.
\end{equation}
The $R$-invariant K\"ahler potential is given by
\begin{equation}
    K(S,\Psi,\bar{\Psi}) = |S|^{2} + |\Psi|^{2} + |\bar{\Psi}|^{2}
    -\frac{\zeta}{4}|S|^{4} + \cdots ,
\end{equation}
where $\zeta$ is a constant of order $1$ and the ellipsis denotes
higher-order terms, which we neglect in the present analysis. We gauge
the U$(1)$ phase rotation:$\Psi \rightarrow e^{i\delta}\Psi$ and
$\bar\Psi \rightarrow e^{-i\delta}\bar\Psi$. To satisfy the $D$-term
flatness condition we take always $\Psi = \bar\Psi$ in our analysis.

As is shown in Ref.\cite{hybrid} the real part of $S(x)$ is identified
with the inflaton field $\sigma/\sqrt{2}$.  The potential is minimized
at $\Psi = \bar{\Psi} = 0$ for $\sigma$ larger than $\sigma_{c}=
\sqrt{2}\mu/\sqrt{\lambda}$ and inflation occurs for $0< \zeta <1$ and
$ \sigma_{c} \lesssim \sigma \lesssim 1$.

In a region of relatively small $\sigma$ ($\sigma_{c}\lesssim \sigma
\lesssim \lambda /\sqrt{8\pi^{2}\zeta}$) radiative corrections are
important for the inflation dynamics as shown in Ref.~\cite{Dvali}.
Including one-loop corrections, the potential for the inflaton
$\sigma$ is given by
\begin{eqnarray}
  V &\simeq& \mu^4 + \frac{1}{2} \zeta \mu^4 \sigma^2 + \frac{4\zeta^2
    +7\zeta +2}{16} \mu^4 \sigma^4\nonumber\\ 
  && +\frac{\lambda^{2}}{128\pi^2} \left [ \left( \lambda \sigma^2 -
      2\mu^2 \right)^2 \ln \frac{\lambda \sigma^2 - 2\mu^2}{\Lambda^2}
    \left( \lambda \sigma^2 + 2\mu^2 \right)^2 \ln \frac{\lambda
      \sigma^2 + 2\mu^2}{\Lambda^2} -2 \lambda^2 \sigma^4 \ln
    \frac{\lambda \sigma^2}{\Lambda^2} \right] ,
\end{eqnarray}
where $\Lambda$ is a renormalization scale. The Hubble parameter
$H_{\rm pre}$ and $e$-fold number $N_{\rm pre}$ are given by
\begin{equation}
    H_{\rm pre} \simeq \frac{\mu^{2}}{\sqrt{3}}
\end{equation}
and
\begin{equation}
        N_{\rm pre} = \left| \int^{\sigma_{c}}_{\sigma_{N_{\rm pre}}}
        \frac{V}{V'} d\sigma \right|,
\end{equation}
where $\sigma_{N_{\rm pre}}$ is the value of the inflaton field
$\sigma$ corresponding to an $e$-fold number $N_{\rm pre}$.

If we define $N_{\rm COBE}$ as the $e$-fold number corresponding to the
COBE scale, the COBE normalization leads to a condition for the
inflaton potential:
\begin{equation}
  \label{eq:COBE-cond}
  \frac{V^{3/2}(\sigma_{\rm COBE})}{|V'(\sigma_{\rm COBE})|}
  \simeq 5.3\times 10^{-4},
\end{equation}
where $\sigma_{\rm COBE} \equiv \sigma_{N_{\rm COBE}}$. In a hybrid
inflation model, density fluctuation is almost scale free, 
\begin{equation}
  n_{\rm pre} = 1+2\left( \frac{V''}{V} \right) -3\left ( \frac{V'}{V}
  \right)^2 \simeq 1,
\end{equation}
where $n_{\rm pre}$ is a spectral index for a power spectrum of
density fluctuations.

\subsection{New inflation}

Now, we consider a new inflation model.  We adopt the new inflation
model proposed in Ref.~\cite{dynamical-tuning}.  The inflaton
superfield $\phi(x, \theta)$ is assumed to have an $R$ charge
$2/(n+1)$ and U$(1)_{R}$ is dynamically broken down to a discrete
$Z_{2nR}$ at a scale $v$, which generates an effective
superpotential~\cite{dynamical-tuning,Izawa-New-inflation}:
\begin{equation}
        W(\phi) = v^{2}\phi - \frac{g}{n+1}\phi^{n+1}.
        \label{sup-pot2}
\end{equation}
The $R$-invariant effective K\"ahler potential is given by
\begin{equation}
    \label{new-kpot}
    K(\phi,\chi) = |\phi|^2 +\frac{\kappa}{4}|\phi|^4 
    + \cdots ,
\end{equation}
where $\kappa$ is a constant of order $1$. 

The effective potential $V(\phi)$ for a scalar component of the
superfield $\phi(x,\theta)$ in supergravity is obtained from the above
superpotential (\ref{sup-pot2}) and the K\"ahler potential
(\ref{new-kpot}) as
\begin{equation}
        V = e^{K(\phi)}\left\{ \left(\frac{\partial^2 
        K}{\partial\phi\partial\phi^{*}}\right)^{-1}|D_{\phi}W|^{2}
        - 3 |W|^{2}\right\},
        \label{eq:new-pot}
\label{SUGRA-general-potential}
\end{equation}
with 
\begin{equation}
        D_{\phi}W = \frac{\partial W}{\partial \phi} 
        + \frac{\partial K}{\partial \phi}W.
        \label{eq:DW}
\end{equation}
This potential yields a vacuum
\begin{equation}
    \langle \phi \rangle  \simeq  
    \left(\frac{v^2}{g}\right)^{1/n}.
\end{equation}
In the true vacuum we have negative energy as
\begin{equation}
        \langle V \rangle \simeq -3 e^{\langle K \rangle}
        |\langle W \rangle |^{2}
        \simeq -3 \left( \frac{n}{n+1}\right)^{2}|v|^{4}
        |\langle \phi \rangle|^{2}.
        \label{vacuum}
\end{equation}
The negative vacuum energy (\ref{vacuum}) is assumed to be canceled
out by a supersymmetry- (SUSY)-breaking
effect~\cite{Izawa-New-inflation} which gives a positive contribution
$\Lambda^{4}_{\rm SUSY}$ to the vacuum energy. Thus, we have a
relation between $v$ and the gravitino mass $m_{3/2}$:
\begin{equation}
  m_{3/2} \simeq \frac{\Lambda^{2}_{\rm SUSY}}{\sqrt{3}} =
  \left(\frac{n}{n+1}\right) |v|^{2}
  \left|\frac{v^{2}}{g}\right|^{1/n}.
        \label{gravitino-mass}
\end{equation}

The inflaton $\phi$ has a mass $m_{\phi}$ in the vacuum with (for
details, see Ref.~\cite{Izawa-New-inflation})
\begin{equation}
  m_{\phi} \simeq n |g|^{1/n}|v|^{2-2/n}.
        \label{inftaton-mass}
\end{equation}
The inflaton $\phi$ may decay into ordinary particles through
gravitationally suppressed interactions, which yields reheating
temperature $T_R$ given by\footnote{
  The decay rate of the inflaton $\phi$ is discussed in
  Ref.\cite{dynamical-tuning}
}
\begin{equation}
    \label{reheat-temp}
    T_R \simeq 0.1 m_{\phi}^{3/2} \simeq 0.1n^{3/2}
    |g|^{3/2n}|v|^{3-3/n}.
\end{equation}
If we take $n=4$ and $g=1$,
\begin{equation}
T_R \simeq 0.8|v|^{9/4} = 1.9\times 10^{18}{\rm GeV} \left
    ( \frac{v}{M_G} \right)^{9/4}.
\end{equation}
In this case, the reheating temperature $T_R$ is as low as $2$~GeV -
$6 \times 10^{4}$~GeV for $v \simeq 10^{-8} - 10^{-6}$ ($m_{3/2}
\simeq 0.02~ {\rm GeV}-2$~TeV), for example, which is low enough to
solve the gravitino problem.

Let us discuss dynamics of the new inflation model.  Identifying the
inflaton field $\varphi(x)/\sqrt{2}$ with the real part of the field
$\phi(x)$, we obtain a potential of the inflaton for $\varphi < v$
from Eq.~(\ref{eq:new-pot}):
\begin{equation}
    \label{new-eff-pot2}
    V(\varphi) \simeq v^4 - \frac{\kappa}{2}v^4\varphi^2
    -\frac{g}{2^{\frac{n}{2}-1}}v^2\varphi^n 
    + \frac{g^2}{2^n}\varphi^{2n}.
\end{equation}
It has been shown in Ref.~\cite{Izawa-New-inflation} that the slow-roll condition
for the inflation is satisfied for $0< \kappa < 1$ and $\varphi
\lesssim \varphi_f$ where
\begin{equation}
    \label{new-inflaton-final}
    \varphi_f \simeq \sqrt{2}
    \left(\frac{(1-\kappa)v^2}{gn(n-1)}\right)^{1/(n-2)}.
\end{equation}
New inflation ends when $\varphi$ becomes larger than $\varphi_f$.
The Hubble parameter of the new inflation model is given by
\begin{equation}
    \label{new-hubble}
    H_{\rm new} \simeq \frac{v^2}{\sqrt{3}}.
\end{equation}
The $e$-fold number $N_{\rm new}$ is given by
\begin{equation}
  N_{\rm new} = \left| \int^{\varphi_f}_{\varphi_{N_{\rm new}}}
    \frac{V}{V'} d\varphi \right| .
\label{eq:N-efold2}
\end{equation}

The amplitude of primordial density fluctuations $\delta \rho/\rho$
due to the new inflation model is written as
\begin{equation}
    \label{eq:new-density}
    \frac{\delta\rho}{\rho} \simeq \frac{1}{5\sqrt{3}\pi}
    \frac{V^{3/2}(\varphi_{N_{\rm new}})}{|V'(\varphi_{N_{\rm new}})|}
    \simeq \frac{1}{5\sqrt{3}\pi} \frac{v^{2}}{\kappa\varphi_{N_{\rm
          new}}}.
\end{equation}
Notice here that we have larger density fluctuations for smaller
$\varphi_{N_{\rm new}}$ and hence the largest amplitude of the
fluctuations is given at the beginning of new inflation. An
interesting point on the above density fluctuations is that it results
in a tilted spectrum with spectral index $n_{\rm new}$ given by (see
Refs.~\cite{dynamical-tuning,Izawa-New-inflation})
\begin{equation}
    \label{eq:new-index}
    n_{\rm new} \simeq 1 - 2 \kappa.
\end{equation}

\subsection{Initial value and fluctuations of $\varphi$}

The crucial point observed in Ref.~\cite{dynamical-tuning} is that
preinflation sets dynamically the initial condition for new inflation.
The inflaton field $\varphi(x)$ for new inflation gets an effective
mass $\sim \mu^2$ from the $e^{K}[\cdots]$ term in the potential
(\ref{SUGRA-general-potential}) during
preinflation~\cite{Hybrid-inflation,Kumekawa}.  Thus, we write the
effective mass $m_{\rm eff}$ as
\begin{equation}
    \label{eff-mass}
    m_{\rm eff} = c\mu^2 = \sqrt{3} c H,
\end{equation}
where we introduce a free parameter $c$ since the precise value of the
effective mass depends on the details of the K\"ahler potential. For
example, if the K\"ahler potential contains $-f|\phi|^2|S|^2$, the
effective mass is equal to $\sqrt{1+f}\mu^2$.  

The evolution of the inflaton $\varphi$ for the new inflation model is
described as
\begin{equation}
  \label{eq:osc-pre}
  \ddot{\varphi} + 3H \dot{\varphi} + m_{\rm eff}^2 \varphi = 0.
\end{equation}
Using $\dot{H}\simeq 0$, we get a solution to the above equation as
\begin{equation}
  \label{eq:sol-pre}
  \varphi \propto a^{-3/2 + \sqrt{9/4 - 3c^2}},
\end{equation}
where $a$ denotes the scale factor of the universe.  Thus, for $c
\gtrsim \sqrt{3}/2$, $\varphi$ oscillates during the preinflation and
its amplitude decreases as $a^{-3/2}$.  Thus, at the end of
preinflation the $\varphi$ takes a value
\begin{equation}
  \varphi \simeq \varphi_{\rm min} + \left( \varphi_i - \varphi_{\rm
      min} \right)\exp\left(-\frac{3}{2}N_{\rm pre,tot}\right),
    \label{coherent}
\end{equation}
where $\varphi_{i}$ is the value of $\varphi$ at the beginning of
preinflation, $\varphi_{\rm min}$ is the value of $\varphi$ at which
the potential has a minimum, and $N_{\rm pre,tot}$ is the total
$e$-fold number of preinflation.

The $\varphi_{\rm min}$ deviates from zero through the effect of the
$|D_{S}W|^2 + |D_{\phi}W|^2 -3|W|^2$ term and the potential has a
minimum~\cite{dynamical-tuning} at
\begin{equation}
    \label{deviation}
    \varphi_{\rm min} \simeq -\frac{\sqrt{2}}{c^{2}\sqrt{\lambda}}
    v\left(\frac{v}{\mu}\right).
\end{equation}
Thus, at the end of preinflation the $\varphi$ settles down to this
$\varphi_{\rm min}$.

After preinflation, the $\sigma$ and $\Psi (\bar\Psi)$ start to
oscillate and the universe becomes matter dominated.  $\Psi$ and
$\bar\Psi$ couple to the U$(1)$ gauge multiplets and decay immediately
to gauge fields if energetically allowed. We assume that masses for
the gauge fields are larger than those of $\Psi$ and $\bar\Psi$. We
also assume that the supersymmetric (SUSY) standard model particles do
not couple to the gauge multiplets. Thus, $S$, $\Psi$, and $\bar\Psi$
decay into light particles only through gravitationally suppressed
interactions and the coherent oscillations of $S$, $\Psi$, and
$\bar\Psi$ fields continue until new inflation starts.  In this period
of the coherent oscillations the average potential energy of the
scalar fields is the half of the total energy of the universe and
hence the effective mass of $\varphi$ is given by
\begin{equation}
  \label{eq:effective-mass}
  m_{\rm eff}^2  \simeq \frac{3}{2} H^2.
\end{equation}
Here and hereafter, we take $c=1$.  The evolution of $\varphi$ is
described by Eq.~(\ref{eq:osc-pre}).  Taking into account $\dot{H} =
(3/2)H^2$, one can find that the amplitude of $\varphi$ decreases as
$a^{-3/4}$.  After the preinflation ends, the superpotential for the
inflaton of the preinflation vanishes and hence the potential for
$\varphi$ has a minimum at $\varphi \simeq 0$.

During the matter-dominated era between two inflations, the energy
density scales as $\propto a^{-3}$, and it is $\mu^4$ and $v^4$ for a
hybrid inflation and a new inflation, respectively, the scale factor
increases by a factor $(\mu/v)^{4/3}$ during this era.  Thus, the mean
initial value $\varphi_b$ of $\varphi$ at the
beginning of new inflation is written as\footnote{
  Here we have assumed that when $\varphi$ begins oscillating just
  after the preinflation, the time derivative of it vanishes.}
\begin{equation}
    \label{eq:init-new-inflaton}
    \varphi_b \simeq \frac{\sqrt{2}}{\sqrt{\lambda}}
    v\left(\frac{v}{\mu}\right)^2 \left[ \cos \left (
        \sqrt{\frac{5}{3}} \ln \frac{\mu}{v} \right) +
      \sqrt{\frac{3}{5}} \sin \left( \sqrt{\frac{5}{3}} \ln
        \frac{\mu}{v} \right) \right].
\end{equation}

We now discuss quantum effects during preinflation.  It is known that
in a de Sitter universe massless fields have quantum fluctuations
whose amplitudes are given by $H/(2\pi)$.  However, the quantum
fluctuations for $\varphi$ are strongly suppressed~\cite{Enquvist} in
the present model since the mass of $\varphi$ is larger than the
Hubble parameter until the start of new inflation.

Let us consider the amplitude of fluctuations with comoving wave
number $k_{b}$ corresponding to the horizon scale at the beginning of
new inflation. These fluctuations are induced during preinflation and
its amplitude at horizon crossing [$k_{b} = a(t_{h})H_{\rm pre}$,
where $t_h$ is a time of horizon crossing] is given by $H_{\rm
  pre}/(2\pi)(H_{\rm pre}/m_{\rm eff})^{1/2}$.\footnote{
This is valid when $m_{\rm eff}$ is greater than $3H_{\rm pre}/2$.}

Since those fluctuations reenter the horizon at the beginning of new
inflation ($t=t_b$), the scale factor of the universe increases from
$t_{h}$ to $t_b$ by a factor of $(H_{\rm pre}/H_{\rm new})=
(\mu/v)^2$.  As we have seen above, the amplitude of fluctuations
decreases as $a^{-3/2}$ during preinflation and $a^{-3/4}$ during the
matter-dominated era between two inflations. Since the scale factor
increases by $(\mu/v)^{4/3}$ during the matter-dominated era and by
$(\mu/v)^2$ from $t_h$ to $t_b$, respectively, it increases by
$(\mu/v)^{2/3}$ from $t_h$ to the beginning of the matter-dominated
era. Therefore, the amplitude of fluctuations with comoving wavelength
corresponding to the horizon scale at the beginning of new inflation
is now given by
\begin{equation}
  \delta \varphi \simeq \frac{H_{\rm pre}}{2\pi} \left(\frac{H_{\rm
        pre}}{m_{\rm eff}}\right)^{1/2}
  \left[ \left( \frac{\mu}{v} \right)^{2/3} \right]^{-3/2} \left [
    \left( \frac{\mu}{v} \right)^{4/3} \right]^{-3/4}
  \simeq \frac{H_{\rm pre}}{3^{1/4}2\pi}
     \left(\frac{v}{\mu}\right)^{2}.  
\label{eq:q-fluctuation}
\end{equation}
The fluctuations given by Eq.~(\ref{eq:q-fluctuation}) are a little
less than newly induced fluctuations at the beginning of new inflation
[$\simeq v^2/(2\pi\sqrt{3}$)]. Moreover, the fluctuations produced
during preinflation are more suppressed for smaller wavelength. Thus,
we assume that the fluctuations of $\varphi$ induced in preinflation
can be neglected when we estimate the fluctuations during new
inflation.

Here let us estimate the $e$-fold number which corresponds to our
current horizon. From Eq.(\ref{reheat-temp}), the reheating
temperature after new inflation is given as
\begin{equation}
    \label{reheat-temp2}
    T_R \simeq 0.1 m_{\phi}^{3/2} \simeq 0.1n^{3/2}
    |g|^{3/2n}|v|^{3-3/n} \simeq 0.8|v|^{9/4}.
\end{equation}
Here and hereafter we take $n=4$ and $g=1$ for simplicity.  The
$e$-fold number is given by~\cite{Liddle-Lyth}
\begin{equation}
  N_{\rm tot} = 62 - \ln\frac{k}{a_0H_0} -\ln \frac{10^{16}{\rm
      GeV}}{V^{1/4}} +\ln\frac{V^{1/4}}{V_{\rm end}^{1/4}} -
  \frac{1}{3} \ln \frac{V_{\rm end}^{1/4}}{\rho_{\rm reh}^{1/4}},
\end{equation}
where $V$ is a potential energy when a given scale $k$ leaves the
horizon, $V_{\rm end}$ is when the inflation ends, and $\rho_{\rm
  reh}$ is the energy density at the time of reheating. Now we can
take $V\simeq V_{\rm end}$, and $\rho_{\rm reh}^{1/4} \simeq {\rm a\ 
  few}\times T_{\rm reh}$. Therefore, for $k=a_0H_0$ (i.e., present
horizon scale),
\begin{equation}
N_{\rm tot} \simeq 67.8 + \frac{17}{12} \ln v.
\end{equation}
Since preinflation lasts after the scale $k_b$ crosses the horizon by
an $e$-folding number $\ln (\mu/v)^{2/3}= (2/3) \ln(\mu/v)$ as
mentioned above, the $e$-folding number corresponding to the COBE
scale can be expressed as
\begin{eqnarray}
  N_{\rm COBE} &=& N_{\rm tot} -N_{\rm new} + \frac{2}{3} \ln
  \frac{\mu}{v}\nonumber\\ 
  &\simeq& 67.8 +\frac{17}{12}\ln v -N_{\rm new} + \frac{2}{3} \ln
\frac{\mu}{v},
\end{eqnarray}
when we consider COBE normalization, Eq.~(\ref{eq:COBE-cond}), we have 
to use this quantity.

Finally, we make a comment on the domain-wall problem in the double
inflation model. Since the potential of the inflaton $\phi$ has a
discrete symmetry [see Eqs.~(\ref{sup-pot2}) and (\ref{new-kpot})],
domain walls are produced if the phases of $\phi$ are spatially
random. However, preinflation makes the phase of $\phi$ homogeneous
with the help of the interactions between two inflaton fields $S$ and
$\phi$ [see Eq.~(\ref{deviation})]. Therefore, the domain-wall problem
does not exist in the present model.

\subsection{Numerical results}

We estimate density fluctuations in the double inflation model
numerically by calculating the evolution of $\varphi$ and $\sigma$.
For simplicity, we take $\zeta =0$.

\begin{figure}[htbp]
    \centerline{\psfig{figure=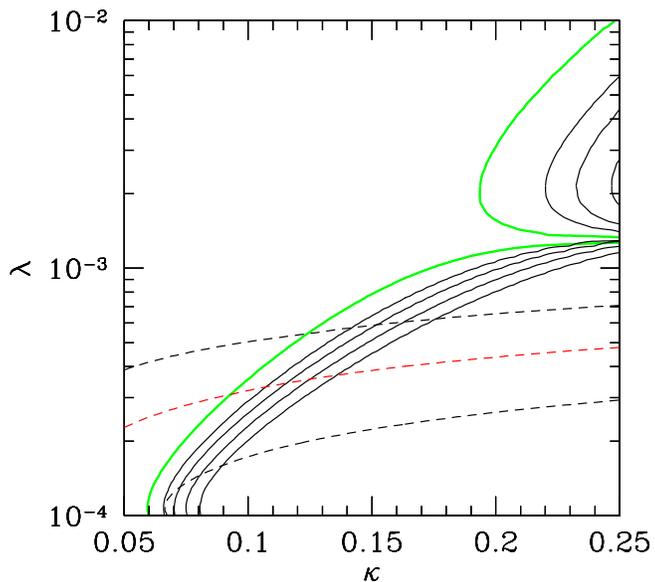,width=9cm}}
    \caption{A double inflation result for $v=10^{-7}$. The solid lines
      correspond to the breaking scale $k_b= 10^{-3}, 10^{-2},
      10^{-1}$ and $1 h$Mpc$^{-1}$, from left to right. Three dashed
      lines represent power spectra ratio ${\cal R} =10, 1$, and
      $0.1$, from top to bottom. The region on the left hand side of
      the thick solid line is irrelevant since COBE scale fluctuations
      are produced during new inflation.}
    \label{fig:inflation_sample}
\end{figure}

Since we are concerned with the situation where the breaking (transit
scale from the hybrid inflation to the new inflation) occurs at
cosmological scale, we choose a parameter region in which the breaking
scale comes within the range
\begin{equation}
  10^{-3}h{\rm Mpc^{-1}} \lesssim k_b \lesssim 1 h{\rm Mpc^{-1}},
\end{equation}
where $h = H_0/(100$kms$^{-1}$Mpc$^{-1})$ and it takes $h=0.5$ in this
paper. Also, we require that the ratio between the density fluctuation
produced by a hybrid inflation and that by new inflation is
\begin{equation}
  0.1 \lesssim {\cal R}\equiv \frac{P_{\rm new}}{P_{\rm pre}} \lesssim
  10 ,
\end{equation}
where $P_{\rm new}$ and $P_{\rm pre}$ refer to the amplitude of a
power spectrum of density fluctuations at $k_b$, produced by new
inflation and preinflation, respectively:

\begin{equation}
P(k) = \left\{
\begin{array}{l}
  P_{\rm pre} \left( \frac{k}{k_b}\right)^1 T^2(k)\quad (k<k_b),\\ 
  P_{\rm new} \left( \frac{k}{k_b}\right)^{n_{\rm new}} T^2(k) =
  P_{\rm pre}{\cal R} \left( \frac{k}{k_b}\right)^{n_{\rm new}}
  T^2(k)\quad (k>k_b),
\end{array}
\right.
\label{break-pk}
\end{equation}
where $T(k)$ is a CDM transfer function. We draw a sample of results
in Fig.\ref{fig:inflation_sample} for $v=10^{-7}$. From this figure we
can see that if
\begin{equation}
  \lambda\sim {\cal O}(10^{-4}-10^{-3})\quad {\rm and}\quad 0.1
  \lesssim \kappa \lesssim 0.2 ,
\end{equation}
$k_b$ is at a cosmological scale, and density fluctuations produced
during new inflation are not too far from that of preinflation.  We
can understand the qualitative dependence of $(k_b, {\cal R})$ on
$(\kappa, \lambda)$ as follows: When $\kappa$ is large, the slope of
the potential for new inflation is too steep, and new inflation cannot
last for a long time. Therefore, the break occurs at smaller scales.
As for ${\cal R}$, we can see that the larger $\lambda$ is, the larger
$\mu$ is, from Eq.(\ref{eq:COBE-cond}). In addition, from
Eqs.(\ref{eq:new-density}) and (\ref{eq:init-new-inflaton}), we can
see that
\begin{equation}
  \left( \frac{\delta \rho}{\rho} \right)_{\rm new} \propto
  \frac{1}{\kappa \varphi_{\rm N_{\rm new}}} \sim \frac{\sqrt{\lambda}
    \mu^2}{\kappa},
\end{equation}
for a fixed $v$. Thus, we have larger ${\cal R}$ for larger $\lambda$.

\section{Comparison with observations}

In this section we compare the result of our double inflation model
with the observations of the cluster
abundances~\cite{Eke,Viana-Liddle} and galaxy
distributions~\cite{da-Costa,Lin,Fisher,Feldman,Tadros}.

\subsection{Cluster Abundances}


Since the power spectrum of density fluctuations shows a break on the
cosmological scale in this double inflation model, we cannot simply
employ the value of $\sigma_8$ quoted by previous
works~\cite{Eke,Viana-Liddle}.  We need to calculate the cluster
abundances by using the Press-Schechter theory~\cite{Press-Schechter}.
According to the Press-Schechter theory, the comoving number density
of collapsed systems of mass $M$ at redshift $z$, per interval $dM$,
is expressed as
\begin{equation}
  \frac{dn(M,z)}{dM} = \sqrt{\frac{2}{\pi}} \frac{\rho}{M}
  \frac{\delta_c(z)}{\sigma^2(M)} \left| \frac{d\sigma(M)}{dM}
    \right| \exp \left[ -\frac{\delta_c^2(z)}{2\sigma^2(M)} \right],
\label{eq:PS}
\end{equation}
where $\rho$ is the mean mass density of the universe at present, and
$\sigma(M)$ is the mass variance, the rms density fluctuations
smoothed over the mass scale $M$, which is defined as
\begin{equation}
  \sigma^2(M) \equiv \frac{1}{(2\pi)^3} \int P(k; A) W^2(kr_0) d^3k,
\end{equation}
where $M= 4\pi {r_{0}}^3 \rho/3$, $W(kr_0)$ is a window function
\begin{equation}
  W(kr_0) = \frac{3}{(kr_0)^3} \left[\frac{}{} \sin (kr_0) - kr_0 \cos
    (kr_0) \right] \frac{}{},
\end{equation}
and $P(k; A)$ is a present matter density fluctuation power spectrum
with a normalization $A$.  For the case of $\Omega_0=1$, $\delta_c(z)=
1.686(1+z)$. This is the density contrast that a collapsed region
should have at collapse time if it had always evolved according to
linear theory. In this paper we take total matter density
$\Omega_{0}=1$ and use the observations of neighbor clusters ($z\simeq
0$).


Given the power spectrum, we can obtain the cluster abundance from the
Press-Schechter theory. When we determine the breaking scale $k_b$,
the power spectrum ratio ${\cal R}\equiv P_{\rm new}/P_{\rm pre}$, and
the spectral index for new inflation $n_{\rm new}$, we can get the
power spectrum up to normalization {$A_{\rm cl}$}. Using this power
spectrum we can calculate the mass variance and obtain, from
Eq.(\ref{eq:PS}),
\begin{equation}
  n(>M_{\rm min} ; A_{\rm cl}) = \int^{\infty}_{M_{\rm min}}
  \frac{dn(M)}{dM} dM.
\label{eq:abundance-PS}
\end{equation}

Many clusters of galaxies are observed using x-ray fluxes. Under the
assumption that clusters are hydrostatic, we can obtain the
mass-temperature relations as
\begin{equation}
  T_{\rm gas} = \frac{9.37\ {\rm keV}}{\beta (5X+3)} \left(
    \frac{M}{10^{15}h^{-1} M_\odot} \right)^{2/3} (1+z) \left(
    \frac{\Omega_0}{\Omega(z)} \right)^{1/3} \Delta_c^{1/3},
\label{eq:mass-temperature}
\end{equation}
where $\Delta_c$ is the ratio of the mean density of a cluster to the
critical density at that redshift, $\beta$ is the ratio of specific
galaxy kinetic energy to specific gas thermal energy, and $X$ is the
hydrogen mass fraction. We take $X=0.76$, $\beta=1$, and $\Delta_c =
18\pi^2 \simeq 178$~\cite{Eke}. Then Eq.~(\ref{eq:mass-temperature})
reduces to
\begin{equation}
  T_{\rm gas} \simeq 7.75 \left( \frac{M}{10^{15}h^{-1} M_\odot}
  \right)^{2/3} \ {\rm keV}.
\label{eq:mass-temperature-reduced}
\end{equation}

The observed cluster abundance as a function of x-ray temperature can
be translated into a function of mass using
Eq.~(\ref{eq:mass-temperature}). Accumulating the observations, Henry
and Arnaud \cite{Henry-Arnaud} gave the fitting formula as
\begin{equation}
\frac{dn(T)}{dT} = 1.8
\left\{
  \begin{array}{l}
    +0.8\\
    -0.5
  \end{array}
\right\}
 \times 10^{-3} 
h^3 {\rm Mpc}^{-3}\ {\rm keV}^{-1}\ \left( \frac{kT}{1{\rm keV}}
\right)^{-4.7\pm 0.5}.
\label{eq:temperature-spectrum}
\end{equation}
By integrating Eq.~(\ref{eq:temperature-spectrum}) we obtain
\begin{equation}
  3.1\times 10^{-4} \left( \frac{T_{\rm min}}{\rm keV}
  \right)^{-4.2} \lesssim
  n[>M_{\rm min}=M(T_{\rm min})]
    \lesssim 8.1\times 10^{-4} \left( \frac{T_{\rm min}}{\rm keV}
    \right)^{-3.2},
\label{eq:integrated-temperature-abundance}
\end{equation}
where the unit of cluster abundance is $h^3 {\rm Mpc}^{-3}$.
Henry and Arnaud~\cite{Henry-Arnaud} also gave a table of cluster
observations whose temperatures are larger than $2.5$ keV, which
corresponds to a lower limit $M_{\rm min}=1.8\times 10^{14}
h^{-1}M_{\odot}$ [see Eq.(\ref{eq:mass-temperature-reduced})].
Therefore we have, from
Eq.~(\ref{eq:integrated-temperature-abundance}),
\begin{equation}
  6.6\times 10^{-6} \lesssim n(>1.8 \times 10^{14}
  h^{-1}M_{\odot})\lesssim 4.3\times 10^{-5}.
\label{eq:abundance-obs}
\end{equation}

\begin{figure}[htbp]
    \centerline{\psfig{figure=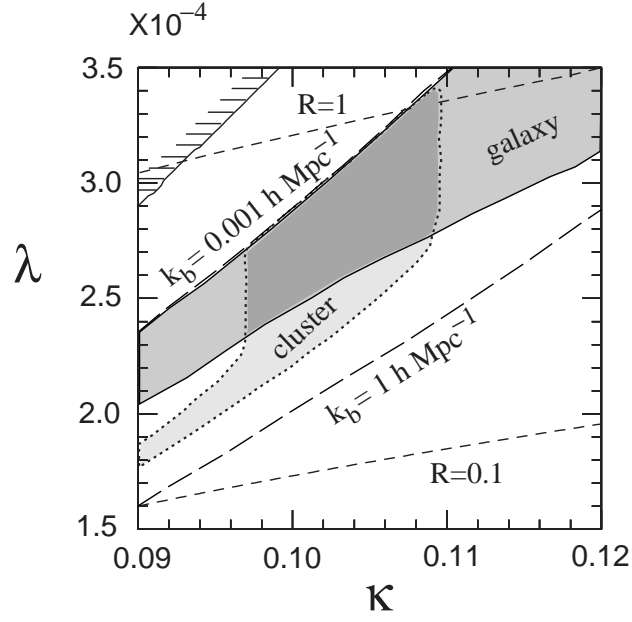,width=9cm}}
    \caption{The allowed parameter region for $v=10^{-7}$. The region
      on a top-left corner is irrelevant since the COBE scale
      fluctuation is produced during new inflation. The lightly shaded
      region inside the dotted line (named ``cluster'') indicates a region
      where ``cluster normalized'' and ``COBE normalized'' $\sigma_8$
      are consistent.  The darkly shaded region satisfies the
      constraints from both the cluster abundances and the galaxy
      distributions ($99\%$ C.L. $\chi^2$ fitting). ${\cal R}$ is a
      power spectra ratio.}
    \label{fig:parameter_v7}
\end{figure}

Matching these abundances, Eq.~(\ref{eq:abundance-PS}) calculated from
the Press-Schechter theory, and Eq.~(\ref{eq:abundance-obs}) inferred
from the x-ray cluster observations, we can determine the
normalization (amplitude) of power spectrum, $A_{\rm cl}$. Using this
normalization, we can obtain ``cluster abundance normalized''
$\sigma_8$, $\sigma_{\rm 8, cl}$, as
\begin{equation} 
  \left.  \sigma_{\rm 8, cl}^2 \equiv \int^{\infty}_{0}
    \frac{k^3}{2\pi^2} P(k; A_{\rm cl}) W^2(kr_0) \frac{dk}{k}
  \right|_{r_0=8h^{-1}{\rm Mpc}}.
\end{equation}
Because of error bars, we have a range of $\sigma_{\rm 8, cl}$ from
observations. On the other hand, we can normalize the power spectrum
by COBE data~\cite{COBE,Bunn-White}.  Therefore, we have ``COBE
normalized'' $\sigma_8$, $\sigma_{\rm 8, COBE}$ together with
$\sigma_{\rm 8, cl}$.  Bunn and White~\cite{Bunn-White} estimates one
standard deviation error of COBE normalization to be $7\%$ which is
much smaller than the one of cluster normalization.  We conclude that
if $\sigma_{\rm 8 ,COBE}$ lies in a $\sigma_{\rm 8, cl}$ range, that
the parameter region of $k_b$, ${\cal R}$, and $n_{\rm new}$ is
consistent with the cluster abundance observations (see lightly shaded
region of Fig.~\ref{fig:parameter_v7} for $v=10^{-7}$).

\subsection{galaxy distributions}

There are many observations which measure the density
fluctuations from galaxy distributions. Among these we use following 
data sets in this paper.
\begin{itemize}
\item Southern Sky Redshift Survey of optically selected galaxies
  (SSRS2) \& The Center for Astrophysics redshift survey of the
  northern hemisphere (CfA2) ($101 {\rm Mpc}/h$ volume-limited, $M_B <
  -19.7 + 5 \log h$), analyzed by da Costa {\it et
    al}~\cite{da-Costa}.
\item The same with above ($130 {\rm Mpc}/h$ volume-limited, $M_B < -
  20.3 + 5 \log h$)~\cite{da-Costa}.
\item The Las Campanas Redshift Survey (LCRS), analyzed by Lin {\it et
    al}~\cite{Lin}.
\item The Infrared Astronomical Satellite (IRAS) 1.2 Jy Sample,
  analyzed by Fisher {\it et al}~\cite{Fisher}.
\item The Queen Mary College, Durham, Oxford, and Toronto (QDOT)
  survey, analyzed by Feldman {\it et al}~\cite{Feldman}.
\item IRAS 1.2 Jy + QDOT [$P(k) =8000$ weighting], analyzed by Tadros
  and Efstathiou~\cite{Tadros}.
\end{itemize}

\begin{figure}[htbp]
    \centerline{\psfig{figure=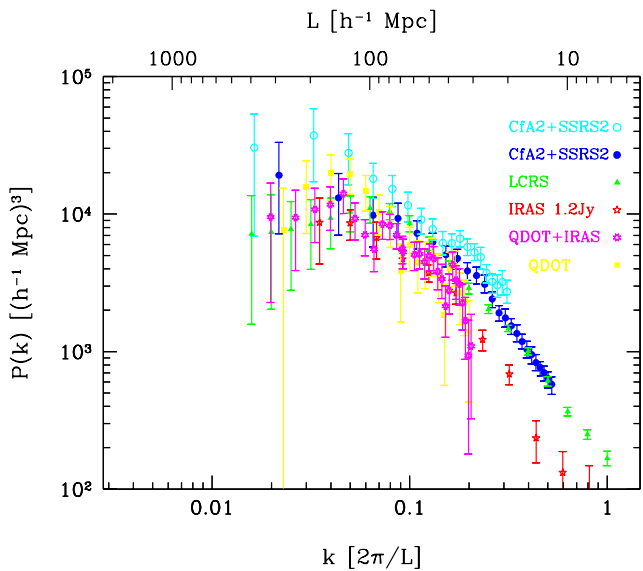,width=9cm}}
    \caption{The observations we have used in this paper to constrain
      our double inflation model. Each symbol represents observations
      (see text, figure courtesy of M. Vogeley).}
    \label{fig:spectra}
\end{figure}

These data are compiled by Vogeley~\cite{Vogeley}. In
Fig.~\ref{fig:spectra}, we plot the observations we use in this paper.

Employing the COBE normalization, we can determine the power spectrum
with its overall amplitude if we fix the the breaking scale $k_b$, the
power spectrum ratio ${\cal R} \equiv P_{\rm new}/P_{\rm pre}$, and
the spectral index for new inflation $n_{\rm new}$.  One might want to
make direct comparison of this power spectrum with above observations
of galaxy distributions.  However, distribution of luminous objects
such as galaxies could differ from underlying mass distribution
because of so-called bias.  There is even no guarantee that each
observational sample has same bias factor.  Therefore, we only
consider the shape of the power spectrum here.  We change the overall
amplitude of each set of observations arbitrarily.  And we estimate
the goodness of fitting by calculating $\chi^{2}$ of this power
spectrum with fixing $k_b, {\cal R}$, and $n_{\rm new}$.


In Fig.~\ref{fig:parameter_v7}, we plot a sample of our results for
$v=10^{-7}$.  There is a parameter region where both the cluster
abundances and galaxy distributions can be accounted for by our model.
What we can see from this figure is that we have almost fixed value of
$\kappa$ and $\lambda$, if we require that a break should occur at a
cosmological scale. The results for $v=10^{-6.5}$ to $v=10^{-7.5}$ are
summarized in Table~\ref{table:couplings} (outside of this range, we
cannot find a suitable parameter region), where we write the coupling
constants $\kappa$ and $\lambda$.

\begin{figure}[htbp]
  \centerline{\psfig{figure=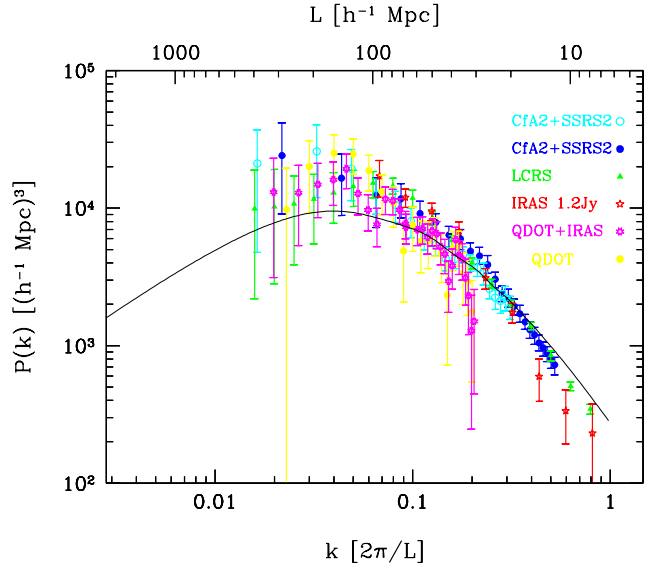,width=9cm}}
    \caption{The power spectrum with optimized galaxy distributions
      for a standard CDM. $\chi^2/{\rm d.o.f.} = 1.74$.}
    \label{fig:ps-standard}
\end{figure}

\begin{figure}[htbp]
  \centerline{\psfig{figure=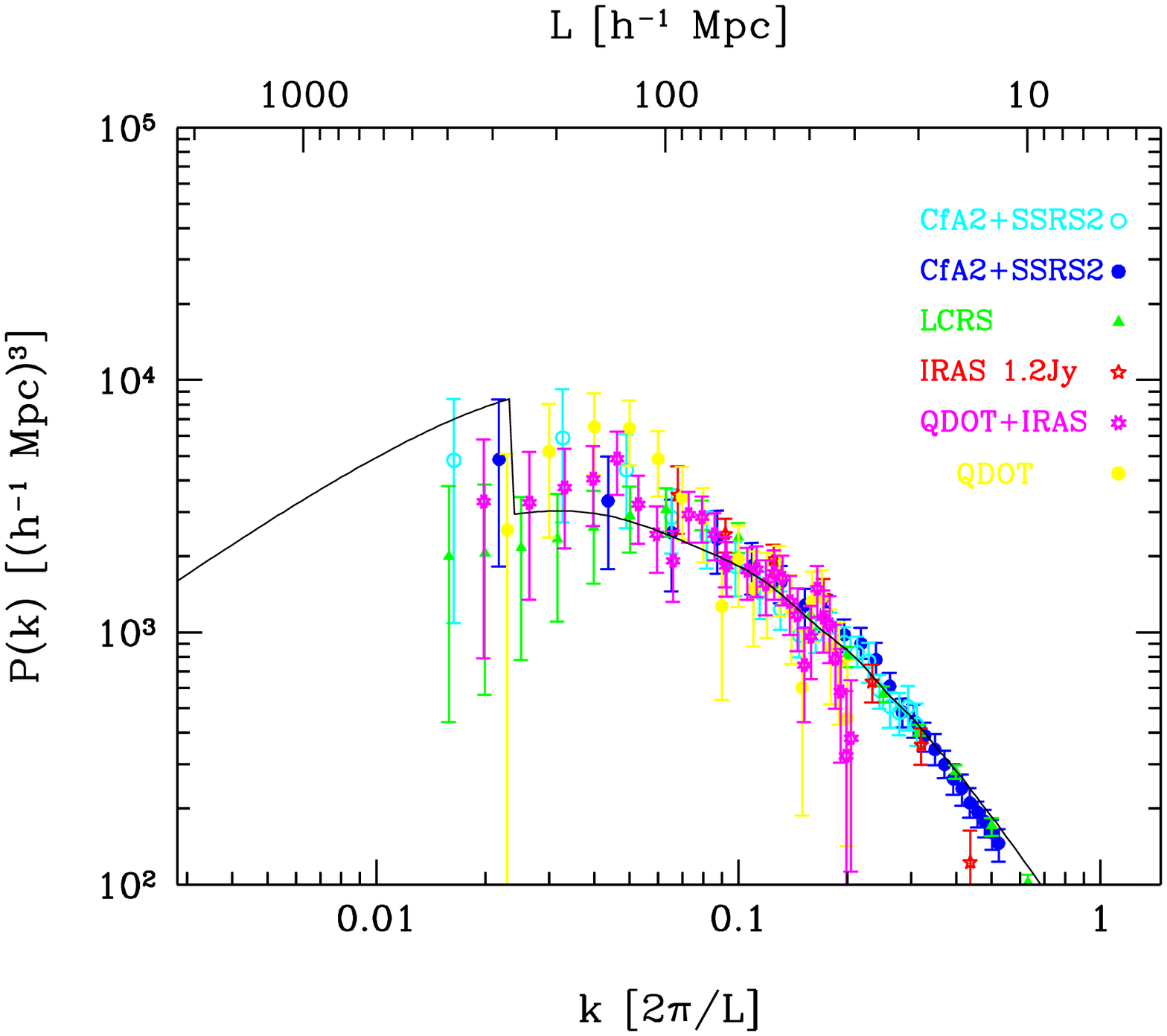,width=9cm}}
    \caption{The power spectrum for $k_b = 0.023746 h^{-1}$Mpc, ${\cal 
        R} =0.345$, and $n_{\rm new} =0.8$. $\chi^2/{\rm d.o.f} =
      1.07$.}
    \label{fig:ps-bestfit}
\end{figure}

In Fig.\ref{fig:ps-standard}, we plot the power spectrum for a
standard CDM model with the optimized galaxy distributions. In this
case, $\chi^2$ normalized by the degree of freedom is $1.74$. Also in
Fig.\ref{fig:ps-bestfit}, we plot the power spectrum for one of the
parameters which minimize $\chi^2 ({\cal R}=0.345, n_{\rm new} =0.8,
k_b = 0.023746h^{-1}$Mpc), where the $\chi^2$ normalized by the d.o.f.
is $1.07$.

\subsection{CMB anisotropies}

\begin{figure}[htbp]
  \centerline{\psfig{figure=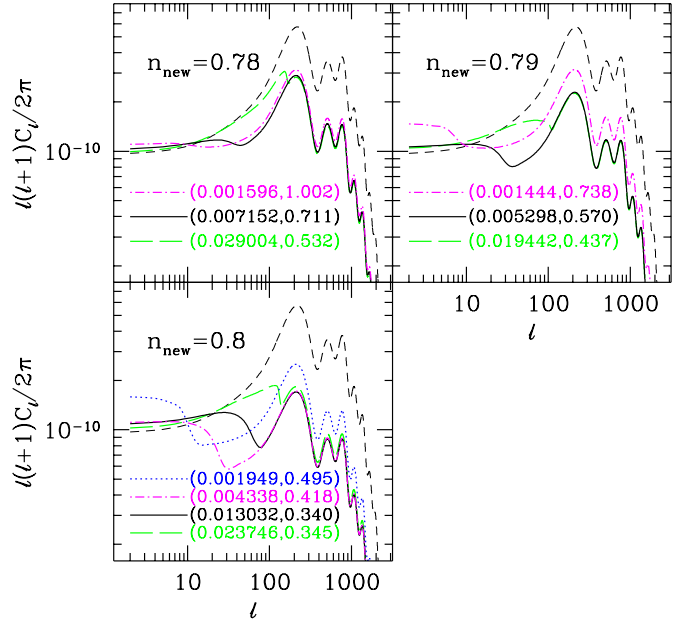,width=9cm}}
    \caption{The CMB angular power spectra for the nine parameter
      sets in Table II. $(k_b, {\cal R})$ with $\Omega_0=1$, $h=0.5$,
      and $\Omega_B=0.06$ (baryon density parameter) are shown in the
      figure.  In a panel for $n_{\rm new}=0.8$, the parameter used in
      the previous figure is also plotted. A standard CDM model with
      same cosmological parameters is plotted in each panel (a dashed
      line) for a reference.}
    \label{fig:C_ell}
\end{figure}

In our double inflation model, the density fluctuations at smaller
scales are produced during new inflation, and they have different
amplitudes from COBE normalization. Also, they are tilted in general
($n_{\rm new} = 1-2\kappa<1$).  Thus, the cosmic microwave background
(CMB) anisotropy angular power spectrum would have a nontrivial shape
at smaller scales. We choose nine parameter sets from
Fig.\ref{fig:parameter_v7} [see Table\ref{table:9prms}], and calculate
the CMB angular power spectra (Fig.\ref{fig:C_ell}).  In the allowed
parameter region, ${\cal R}\lesssim 1$ and $n_{\rm new}<1$.  In this
region, the characteristic feature known as the acoustic peaks is
suppressed compared with the standard CDM case (dashed lines). Some of
them show a dip on the scales, which correspond to the breaking scales
$k_b$, larger than the first peak (smaller in $\ell$).
Although such recent medium angle experiments as
Saskatoon~\cite{Saskatoon}, QMAP~\cite{QMAP}, and
TOCO97/TOCO98~\cite{TOCO} have reported the existence of the first
acoustic peak, these results are inconclusive in view of rather large
observational errors.
The observations of CMB anisotropies by future satellite experiments
[Microwave Anisotropy Probe (MAP)\cite{MAP}, Planck\cite{PLANCK}]
would be able to test our models.

\section{Conclusions and Discussions}

In this paper we have studied the density fluctuations produced in the
double inflation model in supergravity, and compared it to the
observations. Our double inflation model consists of preinflation (=
hybrid inflation ) and new inflation.  Preinflation provides the
density fluctuations observed by COBE and it also dynamically sets the
initial condition of new inflation through supergravity effects.  The
predicted power spectrum has almost a scale-invariant form ($n_{s}
\simeq 1$) on large cosmological scales which is favored for the
structure formation of the universe~\cite{White}.  On the other hand,
new inflation gives the power spectrum which has different amplitude
and shallow slope ($n_{s} < 1$) on small scales.  Thus, this power
spectrum has a break on the scale corresponding to the turning epoch
from preinflation to new inflation.

We have shown that there is a parameter region where the double
inflation model produces an appropriate power spectrum, i.e., the
break occurs at a cosmological scale and both cluster abundances and
galaxy distributions can be accounted for.

We have also calculated the CMB angular power spectra for some
appropriate parameters. In our double inflation model, the acoustic
peaks are suppressed compared with the no-break model. Future
satellite experiments would be able to test our model and will make
precise determination of model parameters possible.

\section*{Acknowledgment}

The authors thank M. Vogeley for kindly providing the data for galaxy
distributions.  T. K. is grateful to K. Sato for his continuous
encouragement, and to T. Kitayama for useful discussions. N.S. thanks
the Max Planck Institute for Astrophysics in Garching for their warm
hospitality.  A part of work is supported by Grant-in-Aid from the
Ministry of Education.

\begin{table}[h]
  \begin{center}
     \caption{$\kappa$ and $\lambda$ for each $v$}
     \label{table:couplings}
     \begin{tabular}{|c||cc|}
        $v$ & $\kappa$  & $\lambda$  \\
        \hline
        $10^{-6.5}$
                & $0.03\sim 0.05$
                & $1.8\times 10^{-4}\sim 2\times 10^{-4}$ \\
        $10^{-6.8}$     
                & $0.075\sim 0.085$ 
                & $1.5\times 10^{-4}\sim 3\times 10^{-4}$ \\
        $10^{-7}$
                & $0.09\sim 0.11$ 
                & $2\times 10^{-4}\sim 3\times 10^{-4}$ \\
        $10^{-7.3}$     
                & $0.125\sim 0.14$ 
                & $2\times 10^{-4}\sim 2.5\times 10^{-4}$ \\
        $10^{-7.5}$     
                & $0.145\sim 0.16$ 
                & $1.5\times 10^{-4}\sim 2\times 10^{-4}$ \\
    \end{tabular}
  \end{center}
\end{table}
\begin{table}[h]
  \begin{center}
     \caption{nine parameter sets.}
     \label{table:9prms}
     \begin{tabular}{|c||cc|cc|}

       $n_{\rm new}$ & $\kappa$ & $\lambda\times 10^4$ &
       $k_b(h^{-1}$Mpc) & {$\cal R$} \\ \hline
        $$
                & $$
                & $2.9$
                & $0.029004$
                & $0.532$\\
        $0.78$
                & $0.109$
                & $3.1$
                & $0.007152$
                & $0.7112$\\
        $$
                & $$
                & $3.344$
                & $0.001596$
                & $01.002$\\
\hline
        $$
                & $$
                & $2.7$
                & $0.019442$
                & $0.437$\\
        $0.79$
                & $0.104$
                & $2.878$
                & $0.005298$
                & $0.570$\\
        $$
                & $$
                & $3.056$
                & $0.001444$
                & $0.738$\\
\hline
        $$
                & $$
                & $2.456$
                & $0.013032$
                & $0.340$\\
        $0.80$
                & $0.098$
                & $2.589$
                & $0.004338$
                & $0.418$\\
        $$
                & $$
                & $2.7$
                & $0.001949$
                & $0.495$\\
    \end{tabular}
  \end{center}
\end{table}


\end{document}